       \documentstyle [12pt]   {article}
 \evensidemargin\oddsidemargin
\begin{document}
\count0 = 1
%
 \title{{ Information Systems Self-description \\
 and Quantum Measurement Problem   \\ }}
\small\author{S.N.Mayburov \\ 
Lebedev Inst. of Physics\\
Leninsky Prospect 53, Moscow, Russia, 117924\\
E-mail :\quad   mayburov@sci.lpi.msk.su}
\date {}
\maketitle
\begin{abstract}

  Information-Theoretical restrictions on the systems self-description
and the information acquisition are  applied to Quantum Measurements Theory.
For the measurement of quantum object S  by  the information system $O$
such restrictions  are  described by restricted states $R_O$
formalism.  
$R_O$ ansatz  can be introduced phenomenologically from 
 the agreement
with Shr$\ddot {o}$dinger dynamics and measurement statistics.
 The analogous  restrictions obtained in Algebraic QM 
 from the consideration of Segal algebra $\cal U_O$  of $O$ observables;
the resulting $O$  restricted states $\{\xi^O_i\}$ set
is defined as $ \cal U_O$  dual space. From
 Segal theorem for associative  (sub)algebras  it's shown
 that  $\xi^O_j$ describes the  random 'pointer'
outcomes $q_j$  observed by $O$ in the individual events.

\end{abstract}
\vspace{6mm}
\small { \quad
To appear in  Int. J. Theor. Phys. (2004)  }  
\vspace{10mm}
%
%
%

\section  {Introduction}

There are several fundamental problems concerning the interpretation of 
 Quantum Mechanics (QM), mainly involving the measurement process
(Jauch,1968; Aharonov,1981). 
 The oldest and most prominent of them is probably
  the State Collapse or Quantum Measurement
   Problem  (D'Espagnat,1990; Busch,1996).
In this paper we shall analyze the quantum  measurements  within the
 Information-Theoretical framework and demonstrate
the importance of such consideration. Indeed,   the
  measurement of any kind results in   the  reception of data about
 the observed system S parameters   by another  system $O$ (Observer) 
 (Guilini,1996; Duvenhage,2002).  
Therefore   the studies of
  Information-Theoretical restrictions on the information which can be
 transferred from  S to $O$  
can be important
in the Measurement Theory (Breuer,1996).
%
In the model regarded here
 $O$ is   the information gaining and utilizing system 
(IGUS);  it processes and memorizes the  information acquired as the result of
 S interactions   with the measuring system (MS) which element is $O$.
 We  assume  that QM description is applicable
both for  a microscopic    and  macroscopic
objects, this is the standard approach in Quantum Measurements Theory
(Busch, 1996). In particular,
 $O$ state supposedly is described 
by the  quantum statistical state  $\rho$ 
  relative to another  observer $O'$ (Rovelli,1995;Bene,2000).
In principle,in our approach $O$  can be  either a human brain
 or some automatic device,  in all cases it's
the system  
  which final  state correlates with the input data.

   S measurement by $O$ can be described by MS state $\rho _{MS}$ evolution
relative to external $O'$, yet  in our approach we regard    
  S information recognition and memorization  performed by $O$
itself, not by $O'$.
 Therefore the reception of information by $O$ during the measurement    
  should be analyzed within the  self-description framework (Svozil,1993).
The  systems self-description
was  studied extensively in 
  Self-reference  Problem context 
 (Finkelstein,1988; Mittelstaedt,1998).
It was shown that the self-description of an arbitrary system 
is always incomplete; this result often interpreted as the
analog of G$\it\ddot{o}$del Theorem for Information theory (Svozil,1993).
 The self-description
 in the measurement process that is called also the  measurement from inside
 was considered in the  formalism of inference maps 
between MS and $O$ (Breuer,1996). It
 leads to some general results for the measurement properties
 but doesn't permit to derive the internal (restricted) $O$
 states $R_O$ from the  first  principles of  the theory.
Following this approach, we propose here the novel
formalism  which under simple assumptions
describes MS states mapping to $O$ states.
In this formalism MS quantum state is represented by 
 the doublet $\Phi= \{ \phi^D, \phi^I \} $, where $\phi^D=\rho_{MS}$ is  MS
density matrix, $\phi^I(n)$ is $O$ restricted state which describes $O$
 subjective information 
in the given individual event $n$  
(Mayburov,2001).
 It follows that $\phi^I$ is 
  the stochastic state which describes the random outcomes
 in the measurement of S states superpositions and this effect
can be interpreted as the subjective state collapse observed by $O$.  
 It will be shown that such formalism
  corresponds to the well-known generalization of standard QM
 - algebraic QM based on Segal and $C^*$- algebras of observables.
 (Emch,1972). In its framework $\phi^D$ is defined on MS 
observables algebra $\cal{U}$,
 $\phi^I$ corresponds to the 
  state defined on $O$ observables subalgebra $\cal{U}_O$.


%
We should stress that  the observer's consciousness 
doesn't play any role in our theory and isn't referred to directly anywhere
(London,1939).
%
 The terms 'perceptions', 'impressions' used here
 are  defined  in strictly physical  terms.
In our model the perception is  
 the acquisition of some information by $O$, i.e. the change
of $O$  state;   a different
 $O$ impressions are associated with a different  $O$ states.
 The systems states are defined in  $O$ reference frame (RF)
 (or other  $O'$ RF)
 and are referred to also as 'S  state for $O$'.


\section {Measurements  and Quantum States Restrictions}

Our formalism exploits both the quantum states in the individual events
- i.e. the  individual states and the 
  statistical states which describe the quantum ensembles properties
 (Mittelstaedt,1998).
 In QM  the individual
states are the pure states  represented by Dirac vectors 
 $|\Psi_l \rangle$    
in $\cal H$ ; the statistical states are  described by  the 
normalized, positive operators of trace $1$
- density matrixes  $\rho$ on $\cal {H}$.
 If   $|\Psi_l \rangle$ composition  is known  for 
the given ensemble, its state can be described in more detail by
 the   ensemble state (Gemenge)    
 represented by the table $W^e=\{ \Psi_l; P_l\}$, 
where $P_l$ are the corresponding probabilities (Busch,1996).
   Algebraic QM   states  will be considered in chap. 3.

We shall start from the analysis of
   simple  MS  model
which    includes
the measured state  S and
 IGUS  $O$   storing the incoming S information.
S represented by two-dimensional state vector $\psi_s$,
whereas $O$ is described by three-dimensional 
%
  Hilbert space $\cal {H}_O$. Its
  basis  consists of 
  the orthogonal states $|O_{0,1,2}\rangle$,
 which are  the eigenstates of $Q_O$  
'internal pointer' observable with eigenvalues $q^O_{0,1,2}$.
 In our model the detector D is omitted in  MS chain,
the role of $O$ decoherence effects will be discussed below.
Let us   consider    
 the measurement   of  S observable $\hat{Q}$
for MS initial state:
\begin {equation} 
     \Psi^{in}_{MS}=\psi_s |O_0\rangle= 
(a_1|s_1\rangle+a_2|s_2\rangle)|O_o\rangle \, , \label {AAB}
\end {equation}
 where $|s_{1,2}\rangle$ are $Q$ eigenstates with eigenvalues $q_{1,2}$.
  S,$O$  interaction $\hat{H}_I$ starts at $t_0$
and finished effectively at some finite $t_1$, in that case    
 Schr$\ddot{o}$dinger equation for $\Psi^{in}_{MS}$ 
supposedly results in
 MS final  state $\rho^p_{MS}$ :
\begin {equation}
   \Psi_{MS}=\sum \Psi^{MS}_j = \sum a_i|s_i\rangle|O_i\rangle \, .
                                   \label {AA2}
\end {equation} 
It turns out that $ \bar{Q}_O=\bar{Q}$, so
 $O$ performs the unbiased $Q$ measurement (Von Neuman,1932).
Meanwhile for any $O$ observable
 $Q'_O  \ne F(Q_O) ;\, \bar{Q}'_O=0$ independently of 
$\psi_s$. 
Concerning  the information recognition by IGUS $O$, 
 we  assume  that $|O_{1,2,0}\rangle$ correspond to 
the certain information patterns - an impressions percepted by $O$
 as $Q_O$ values $q^O_{1,2,0}$ (Guilini,1996).
%
%
For the external observer $O'$  at $t>t_1$, 
MS is  in the pure state  $\Psi_{MS}$ of (\ref {AA2});
whereas from $O$ 'point of view' $\Psi_{MS}$ describes the
simultaneous superposition (coexistence) of two
contradictory impressions : $Q_O=q^O_1$ and $Q_O=q^O_2$ percepted by $O$
 simultaneously.
However, it's well known   that experimentally   
the macroscopic   $O$ observes at random one of $Q_O$ values $q^O_{1,2}$
in any individual event. It means 
that  S final state is $|s_1\rangle$ or $|s_2\rangle$  and
S state collapse occurs. 
   S final  state  is described by
  the density matrix  of mixed state: 
\begin {equation}
 \rho_s^m= \sum_i |a_i|^2|s_i \rangle \langle s_i|
                                                              \label {AA33}
\end {equation}
In accordance with it one
 can  ascribe to MS the mixed state :
\begin {equation}
 \rho_{MS}^m= \sum_i |a_i|^2|s_i \rangle \langle s_i||O_i \rangle \langle O_i|
                                                              \label {AA3}
\end {equation}
which    differs principally from $\rho^p_{MS}$ of (\ref {AA2}). 
  This   discrepancy constitutes the basis of  Wigner 
  'Friend  Paradox' for $O, O'$ (Wigner,1961).
  We shall propose here   the formalism which incorporate consistently
this two   
  MS descriptions 'from outside' by $O'$
and 'from inside' by $O$.

Generally,  the  measurement 
 of an arbitrary  system $S'$ is the 
mapping of $S'$ states set $N_S$ to  the given IGUS $O^S$ states set $N_O$.   
 If the final $O^S$ and $S'$ states can't be factorized,
 then  $O^S$ should be considered as
 the subsystem of the large system $S_T=S'+O^S$ with the states set $N_T$
(Mittelstaedt,1998).
  In this situation - 'measurement from inside',  $N_O$ is $N_T$ subset and 
  the inference map of $S_T$ state  to $N_O$ defines
$O^S$ state  called also  the restricted  state $R_O$.
 The important property of
 $S^T \rightarrow O^S$  inference map is formulated by
 Breuer theorem: if for two arbitrary $S_T$ 
states $\Lambda_{S},\Lambda'_{S}$ 
their restricted  states $R_O, R'_O$ coincide, then for $O^S$ this $S_T$ 
states are indistinguishable (Breuer,1996).
  $O^S$ has less degrees of freedom
  than $S_T$ and 
 can't discriminate all possible $S_T$ states,  due to it
some number of $S_T$ indistinguishable states should exist for any $S_T$
 and $O^s$ (Svozil, 1993).
 In quantum case  
  the  observables noncommutativity and nonlocality introduce
some novel features regarded below.
Despite that $R_O$ are incomplete $S_T$ states,
 they are the real physical states
for $O^S$ observer - 'the states in their own right', as Breuer 
characterizes them. 

 The obtained
$S'$,$O^S$,$S_T$ relations are applicable to our MS model which can be
treated as  MS measurement from inside. 
  Breuer's results leave the considerable freedom for the choice of
the inference map and
 don't permit
to derive the restricted states ansatz 
directly.
 It was proposed phenomenologically (Breuer,1996)  
that $R_O$ coincides with MS  partial trace 
 which for MS final  state  (\ref {AA2}) is equal:
\begin {equation} 
   R_O=Tr_s  {\rho}^p_{MS}=\sum |a_i|^2|O_i\rangle\langle O_i|
      \label {AA4}
\end {equation}
%
%
%
For   MS mixed  state $\rho^{m} _{MS}$
of (\ref {AA3}) 
 the  corresponding restricted 
 state is the same $R^{mix}_O=R_O$.
This equality doesn't mean  the
 collapse of MS pure state $\Psi_{MS}$
 because  the collapse appearance
   should be   verified also  for  MS individual
  states.
For the pure case MS individual state is  $\Psi_{MS}$ of (\ref{AA2}), yet
 for  the incoming S statistical mixture  - Gemenge (\ref {AA33})
 it  differs
from event to event:
\begin {equation}
\varsigma(n)=\rho^I_l=|O_l\rangle \langle O_l|| s_l\rangle\langle s_l|
  \label {A44}
\end {equation}
where the random $l(n)$ frequency is stipulated by the
 probabilistic distribution $P_l=|a_l|^2$.
In any event $\varsigma(n)$  differs from
 MS  state (\ref{AA2})  and 
its restricted state 
 $\varsigma^O(n)=|O_l\rangle \langle O_l|$
  also differs   from $R_O$ of (\ref {AA4}) in any event. Because of it 
 for the
restricted individual states the main condition of Breuer  Theorem
is violated
 and $O$ can differentiate pure/mixed states 'from inside'
in the individual events (Breuer,1996). Therefore the proposed 
formalism doesn't result in 
 the state collapse  for the measurement from inside
 in  standard QM.
Note also that the formulae (\ref {AA4}) 
describes also   the
  restricted $O$   statistical state
$R^{st}_O=R_O$ for the  pure and mixed MS  state $ \rho_{MS}^{p,m}$ 
and this ansatz is correct  both in $O$ and $O'$ RFs.



Note that   in Breuer  theory 
 $O$ can't observe the difference between 
  MS   states with different $D_{12}=a_1^*a_2+a_1 a_2^*$. 
 Such difference revealed by 
  MS interference term (IT) observable:
\begin {equation}
   B=|O_1\rangle \langle O_2||s_1\rangle \langle s_2|+j.c.
    \label {AA5}
\end {equation}
which characterizes $O$,S quantum correlations.
Being measured by external $O'$ on S,$O$,
 it gives $\bar{B}=0$ for the mixed MS state (\ref {AA3}),
 but  $\bar{B}\neq 0$  for the pure MS states (\ref{AA2}).
Yet  $B$ value  can't be measured by $O$ 
'from inside', so $O$,S correlations are 
unavailable for $O$ directly. 
In standard QM $R'_O=R_O$
 is regarded  as $O$ individual (partial) state in $O'$
RF (Lahti,1990). Such $R'_O$ choice is advocated first of all by the correct
$\bar{Q}_O,\bar{Q}'_O$ values obtained for that state.
Let's consider as the example MS state of (\ref {AA2})
 with $a_{1,2}=\frac{1}{\sqrt{2}}$, in this state
$B$ has  the eigenvalue $\tilde {b}=1$.
 It means that in $R'_O$ state  $Q_O$ is prncipally uncertain for $O'$:
 $q_1^O \le \tilde {q}^O \le q^O_2$. In particular,  such $B$ value excludes
the ignorance interpretation of $Q_O$ uncertainty for $O'$,
which assumes that in that state $Q_O$ in any event is sharp and
is equal either to $q^O_1$ or $q^O_2$ (Busch,1996).
However, such reasoning fails for $O$ observer because
$B$ value is unobservable for him, and therefore it's impossible
to conclude in that theory whether $Q_O$ is uncertain or sharp in $O$ RF.   
 Hence $R_O$ of (\ref {AA4}) is just
the phenomenological choice which should be advocated additionally,
and other solutions for $O$ individual  state, as we argue
below, are possible. In particular, in this case the 'subjective'
ignorance interpretation, in which $Q_O$ value is sharp for $O$
but is  simultaneously uncertain for $O'$, can't be excluded.

%
%
%

MS individual state  for $O$ can be rewritten  in doublet 
form $\Phi^B(n)=|\phi^D,\phi^I \gg$ where $\phi^D=\rho_{MS}$
is the dynamical
state component,   the information component
 $\phi^I$ describes $O$ subjective information  in the given event $n$.
 In Breuer theory
 $\phi^I$ is just $\phi^D$  trace, however,
 in the alternative self-description formalism described below it 
 represents the novel $O$ state features.
 As the example of the simple $O$
%
%
%
In this case    
the state collapse 
appears in MS  measurement from inside  and
is described by the information component $\phi^I$ (Mayburov, 2001).
%
To agree with the
 quantum Schr$\rm\ddot{o}$dinger dynamics  (SD),
 this doublet state formalism (DSF) should satisfy
 to two operational conditions : \\
i) if an arbitrary system $S'$   doesn't interact with IGUS $O^S$,
 then for $O^S$ this  system evolves according to
 Schr$\rm\ddot{o}$dinger-Liouville  equation  (SLE)  \\
ii) If $S'$ interacts with $O^S$ and
 the measurement of some $S'$
observable occurs, then  SD can be violated for $O^S$ but,
  as follows from  condition i),   
    $S',O^S$ evolution
 for the external   observer  $O'$ should obey SLE.\\
  Below it will be shown that
DSF  corresponds to the measurements
description in Algebraic QM.
For DSF state   
 $\Phi=|\phi^D, \phi^I\gg$ 
the dynamical component  is also
equal to QM density matrix  $\phi^D=\rho$ and obeys     SLE  :
\begin {equation}
  \frac {\partial \phi^D}{\partial  t} =[\phi^D,\hat {H}]    \label {AA8}
\end {equation}
  For our MS model the initial $\phi^D$ of (\ref{AAB}) evolves   
 at $t>t_1$ to $\phi^D(t)=\rho^p_{MS}$ of  (\ref{AA2}).
   For $t \le t_0$,  the initial
 $\phi^I=|O_0\rangle \langle O_0|$; 
   after S measurement finished
 at $t>t_1$,   its $\phi^I$ outcome  is supposed to be stochastic:
 $\phi^I(n)=\phi^I_i$,  where
$\phi^I_i=|O_i\rangle \langle O_i|$;
 here $i(n)$ 
 described by  the probabilistic distribution with $P_i=|a_i|^2$.
 It turns out  that such doublet individual state $\Phi(n)$ 
can change from event to event,  and $\phi^I(n)$
is partly independent of $\phi^D$ being correlated with it only
 statistically.
  $O$ subjective states $\phi^I$ ensembles coincide for
the pure and mixed states with the same $|a_i|^2$. Therefore the conditions of
 Breuer theorem are fulfilled and  the subjective state
collapse is observed by $O$. 
%

 An arbitrary ensemble evolution can be described
 via the doublet statistical states  
$  |\Theta\gg=|\eta_D,\eta_I\gg  $,
where $\eta_D=\phi^D$, $\eta_I(t)$ describes 
the probabilistic distribution $\{P^f_i(t)\}$  of  $\phi^I_i$ observations
 by $O$ at given $t$:
\begin {equation}
P^f_j(t)=tr(\hat{P}^O_j \eta_D(t)) \label {CC} 
\end {equation}
where $\hat{P}^O_j$ - the projector on the given $\phi^I_j$.
For MS in particular,
 if S don't interact with $O$ (no measurement), 
 then $\eta_I$ is time invariant and MS obeys  the
standard QM SLE evolution for the dynamical component $\eta_D$.
Therefore our doublet states are important only for the
 measurement-like  processes
with direct  $ S-O$ interactions.
%
 $\eta_I(t)$ is defined by $\eta_D(t)$ which  obeys  SLE, 
 because of it
 $\Theta$ evolution is reversible and the acquired $O$ information
can be erased completely (Mayburov,2002).

Plainly, in this theory 
the quantum  states for external $O'$ (and other observers)  also has
the same doublet form $\Phi'$. In the regarded situation
 $O'$ doesn't interact with MS  and so  $O'$ information $\phi^I$
doesn't change during S measurement. Consequently,
 MS  state evolution  for $O'$ described by
$\phi^D$, which obeys SLE. Because of it  MS state
collapse isn't observed by  $O'$ in agreement with  the conditions i, ii.
 As the example of the simple $O$
 toy-model  can be regarded the hydrogen-like atom $A_H$
 for which $O_0$ is  its ground state
 and $O_i$ are the metastable  levels excited by $s_i$, resulting so into the
 final S - $O$ entangled state.
 In our model 'internal pointer' $O_i$
 and $O$ memory  which normally differ supposed to be the same object.
  Witnessing QM Interpretation proposed by Kochen  
  is quite close to DSF but doesn't exploits the self-description approach
(Kochen,1985; Lahti,1990).   

%



%




In DSF $|O_i\rangle $ constitutes the  
 preferred basis (PB) in $\cal H_O$ and its appearance should be explained
in the consistent theory;
this problem called also the basis degeneracy   is well-known
 in  Quantum Measurement Theory  (Lahti,1990; Elby,1994). 
In its essence, the theory consistency demands that the final
 MS states decomposition should be unique,
 but this isn't the case for $\Psi_{MS}$ of (\ref {AA2}).
In DSF  PB problem acquires the additional aspects related
to the information recognition by $O$. 
The plausible explanation prompts $O$ decoherence - i.e.
$O$ interaction with environment E (Zurek,1982).
Such interactionresults in the final entangled S,$O$,E state
which decomposed on some orthogonal $O$ basis  $|O^E_i\rangle$
 (Guilini,1996). 
 Tuning the   $H_{O,E}$ interaction
parameters,   $|O^E_i\rangle$ basis can be made equivalent to 
$|O_i\rangle$ basis. 
For the initial MS state $\Psi^{in}_{MS}$ of
 (\ref {AAB}) it results in the final MS-E state :
\begin {equation}
    \Psi_{MS+E}=\sum a_i|s_i\rangle|O_i\rangle|E_i\rangle
                                             \label {DD1}
\end {equation}
where  $|E_i\rangle$ are final E  states.
It was proved that such triple decomposition is unique, even
if $|E_i\rangle$ aren't orthogonal (Elby,1994).
 It isn't necessary
in our case to use $ N\rightarrow \infty$ limit, E can be also  a finite
 system.

 In addition, the  decoherence results in the
important consequences for  the   information storage by
$\phi^I$.
Really, for the practical IGUS $O$
the memorized states $|O^C_i\rangle$ excited by $|s_i\rangle$ signals
 must be stable or at least long-living. But as follows from eq. (\ref{DD1}),
 any  state $|O^C_j\rangle$ different from one of $|O_i\rangle$
in the short time would split into $|O_i\rangle$ combinations - entangled
$O$,E  states superpositions, so that 
 $|O^C_j\rangle \ne|O_i\rangle$ are in fact a virtual states, 
%
Consequently, in this model  $O$-E decoherence  selects 
 the basis of long-living $O$ eigenstates which supposedly
describes $O$ events perception and
memorization. 
It means that, if our S signal is $Q$ eigenstate which
 induces $Q_O$ eigenstate $|O_i\rangle$, then it's memorized
by $O$ for the long time. It makes also $Q_O$  the preferable observable
for $O$ observables set, since for any
$Q'_O \ne F(Q_O)$ it gives $\bar {Q}'_O=0$ for  $\rho_{MS+E}$
resulting from the arbitrary $\rho_{MS}$ 
 (even if it's not the entangled state of
(\ref {AA2})).  This measurement 
scheme denoted as MS+E model will be considered below,
together with  MS model in which PB settled $a\, priory$.
 In other aspects
 decoherence doesn't change our selfmeasurement model;  its
most important role is the unambigous definition of $O$ PB.
In fact, $\cal {H}_O$ symmetry broken dynamically 
by $H_{O,E}$ interaction which makes majority of $O$ states unstable.
Despite that $O$,E decoherence clearly indicates PB existence,
 it doesn't mean that our DSF follows from first QM principles.
Our theory is still phenomenological and decoherent PB in MS+E
model only reveals
how $\phi^I$ basis should be chosen if such theory is correct.


\section {   Quantum Measurements in Algebraic QM}

Now   the quantum  measurements and $O$ selfdescription
 for the finite quantum systems will be regarded
in  Algebraic QM formalism (Bratelli,1981).
  Besides the standard quantum effects,
Algebraic QM    describes successfully
the phase transitions and other nonperturbative phenomena 
which  standard QM fails to incorporate (Emch,1972). 
Consequently, there are the serious reasons to regard Algebraic QM
as the consistent generalization of standard QM with the fixed Hilbert space. 
Algebraic QM was applied extensively
 to the superselection models of quantum measurements,
in which the detector D or environment E  are regarded as the infinite systems
with $ m,V \rightarrow \infty $ (Primas,1990; Guilini,1996).
The algebraic formalism of nonperturbative QFT  
was used also in  the study of measurement 
dynamics in some realistic  systems (Mayburov,1998).
%
%

In standard QM the fundamental structure is the  states set 
- Hilbert space $\cal H$ on which an observables - Hermitian operators
are defined. It can be shown that for some nonperturbative
systems  the structure of states set   differs principally
 from  $\cal{H}$, and the standard QM
axiomatics becomes preposterous. In distinction, 
   the fundamental structure  of Algebraic QM
 is   Segal algebra $\cal{U}$ of observables
which incorporates the  main properties of  regarded system $S_f$ (Emch, 1972).
 It's more   convenient technically   to  deal with 
 $C^*$-algebra $\cal{C}$  for which $\cal{U}$ is the subset.
   Roughly speaking, $\cal{C}$ is an algebra of
 complex elements for which $\cal{U}$ is the subset 
 of its real (Hermitian) elements. 
  This elements in Algebraic QM
  are the linear operators for  which the sum $A+B$ and
   product $A \cdot B$  defined.
 For our problems $\cal{C}$, $\cal{U}$ are in the unambiguous correspondence: 
$\cal{C} \leftrightarrow \cal{U}$,
and below their use is equivalent in this sense.
   $S_f$   states set $\Omega$ defined by
 $\cal{U}$    via the notorious
 GNS  construction; it proves that
  $\Omega$ is the  vector space dual to
 the corresponding  $\cal C$ (Bratelli,1981).
 Such states are
called here the algebraic states $\varphi $ and
are  the normalized, positive, 
linear functionals on $\cal {U} $: for any observable $A \in \cal{U}$,
  $ \, \forall \varphi \in \Omega;$, 
$\,\bar{A}=\langle \varphi;A\rangle$.
%
%
The  pure states -  i.e.   $\Omega$  extremal points are 
  regarded  as the algebraic individual states
 (AIS) $\xi$; their set denoted $\Omega^p$  (Emch,1972; Primas,1990).
%
 The algebraic mixed states $\varphi_{mix}$
can be constructed as $\xi_i$ ensembles,
  the   ensemble states $W^A$ are
  defined analogously to  their  QM ansatz.
Here only  a finite-dimensional $S_f$ will be considered,
 for  them  $\varphi$ states set is isomorphic to some
 QM density matrixes $\rho$ set. In that case 
we regard the corresponding $\varphi$ and $\rho$ as equivalent,
despite that, strictly speaking, they are  the different 
 objects  (Segal,1947).


  In many practical situations  only some restricted 
   linear subspace $\cal M_R$  or  subalgebra
 $\cal{U}_R$ of $S_f$   algebra $\cal U$
is available for the observation.
 For such subsystems the restricted  algebraic states $\varphi_R$
 and their set $\Omega_R$ can be defined consistently
 via the expectation values of $A_R \in \cal{U}_R$:
\begin {equation}
    \bar{A}_R=\langle \varphi;A_R\rangle=\langle \varphi_R;A_R\rangle
  \label {CC12}
\end {equation}
 $\varphi_R$ doesn't depend on any $A' \notin \cal{U}_R$,
thereby $\forall \varphi_R,\,\langle \varphi_R;A'\rangle=0$.
%
%
Remind that any classical system $S^c$  can be
described by some associative Segal algebra $\cal {U}^C$
 of $S^c$ observables $\{A\}$;
 in algebraic QM $\cal U$ associativity
corresponds to its observables commutativity (Emch,1972).
The theorem by Segal   proves that 
any  associative Segal  (sub)algebra $\cal{U}'$
is isomorphic to some algebra $\cal {U}^C$ of classical  
observables, its  $\varphi^a$ states set
  $\Omega^a$ is isomorphic to
  the set $\Omega ^c$ of the  classical statistical states $\varphi^c$
(Segal,1947).
The corresponding AIS   - i.e. the pure states   
correspond to the classical
individual  states $\xi^c_i$ - points in $S^c$ phase space.
For the systems self-description the most important is the  case when
  $\cal{U}'$ is elementary, i.e. includes only $I$ - unit operator
 and  the single $A \ne I$, 
   then $\xi^c_i=\delta(q^A-q^A_i)$ corresponds to    
 $A$ eigenvalues $q^A_i$ spectra.
 Consequently, even if quantum $S_f$ is described by nonassociative $\cal{U}$,
 it should  contain the subalgebras $\cal{U}'\in U$
   for which                 
 the restricted states are classical.
%



For the classical observing system $S^c_T=S'+O^c$ 
 described by some $\cal{U}^C$,
IGUS  $O^c$  self-description  restrictions are simple and straightforward -
the restricted $S^c_T$ states depend only on those
 $S^c_T$ coordinates $\{x^O_j\}$,
 which are $O$ internal degrees of freedom (Breuer,1996). 
They constitute the subalgebra $\cal{U}^C_R \in \cal{U}^C$.  In practice
 $O^c$ effective subalgebra $\cal{U}^C_O \in \cal {U}^C_R $ which
really defines the measurements 
 can be even smaller, because
some $x^O_j$ can be uninvolved directly into the measurement process.  
QM  Correspondence principle 
   prompts that for  the quantum IGUS $O$
the  restricted $O$ subalgebra $\cal{U}_R$ also should
include only $O$ internal observables. In this case
any effective self-description subalgebra 
 $\cal{U}_O$ belongs to $\cal{U}_R$,
   their  states  sets are denoted $\Omega_O, \Omega_R$ correspondingly.   
 The  main assumption of our theory is as follows:
 given  the subalgebra $\cal{U}_A \in \cal{U}$, 
  in any individual event $n$ an arbitrary  MS AIS $\xi^{MS}(n)$ 
 induces  some    restricted AIS $\xi^A(n)$ spanned on $\cal {U}_A$.
This hypothesis seems to us quite natural  and  
 below the additional arguments in its favor will be presented
for the particular $\cal{U}_A$.
For illustration we regard it first 
for $\cal U_R$, and  accept $ad\;hoc$ that, according to the
formal  definition of the individual states,
such states are $\Omega_R$ extremal points.



  MS is described by
 $ \cal{U}$ Segal algebra  of MS  observables  which defines
  $\varphi^{MS} \in\Omega$ properties,
  $MS+E$ model involves $\cal {U}_{MS,E}$ algebra correspondingly.
 $O$  subalgebra  $\cal{U}_R$  includes $I$ and all $O$ 
internal observables, so it means that
  $\Omega_R$
 is isomorphic to $O$  statistical states $\rho^O$ set $\Gamma_O$.
 Consequently, 
$O$ AIS set $\Omega^p_R$ is isomorphic to $\cal{H}_O$ and 
 any $O$  AIS $\xi^R_i $ corresponds to some  
 state vector $ |O^R_i \rangle \in \cal H_O$.
We don't study here $\xi^R$  states in detail because they
 are unimportant for our aims,
note only that Breuer  state (\ref {AA4}  ) for $O$ $R_O \notin  \Omega^p _R$,
and can't be $O$ AIS  on $\cal {U}_R$ for  an arbitrary $a_{1,2} $. 
%
To define   $\cal U_O$,
 remind that
  for the regarded MS dynamics  $O$ 
 can measure only the  observable $Q_O$. 
For any other $Q'_O \neq F(Q_O)$   one obtains  $\bar{Q}'_O=0$
for any $\Psi^{in}_{MS}$,
 therefore for  any  $\varphi^O \in  \Omega_O$ it  follows that  
 $ \langle  \varphi^O; Q'_O \rangle=0$.
%
%
%
It means that  
  $O$ effective    subalgebra $\cal{U}_O$ is equal to
the elementary $\cal U^I_R$, which
 includes only  $Q_O$ and $I$.
Really, only in this case  
$ \langle   \varphi'; Q'_O \rangle=0$
 for all $ \varphi' \in \Omega^I_R$ defined on  $\cal U^I_R$;
 each $\varphi'$ corresponds to $\varphi^O$ with the same $\bar{Q}_O$
and vice versa;
 so  $\varphi^O$ set $\Omega_O$  is isomorphic to $\Omega^I_R$.
 There is no other $\cal U_R$ subalgebras with such properties and
  that settles $\cal U_O $ finally. It turns out that
 the obtained $\varphi^O$ are  equivalent to $R^{st}_O$ of  (\ref{AA4}) which
 are    MS  statistical states $\rho_{MS}$ restriction to $O$.

%
From Segal theorem     the restricted algebraic $O$ 
 states $\varphi^O\in \Omega_O$   are isomorphic to the classical
 $q^O_i$ distributions;
meanwhile $O$ AIS $\xi^O$ -  $\Omega_O$ extremal points
 are the positive states:
\begin {eqnarray}
            \xi^O_i= \delta(q^O-q^O_i) \label {AA66}
\end {eqnarray}
 which correspond to the classical pointlike  states.
%
  In particular, such $\xi^O_i$ can appear in 
$\psi_s=|s_i\rangle$  measurement  by $O$ as the restriction of MS final
state $\xi^{MS}_i=|s_i\rangle|O_i\rangle$.
%
As was supposed above,  the formal individual state $\xi^O_i$
  percepted by  $O$  as the definite 
$Q_O$ value   $q^O_i$. Such information pattern permits to reveal
 $\xi^O_{i,j}$ distinction in the single event per each state
which is important for our theory interpretation.
Note also that Algebraic formalism, in fact,  extracts PB
 $|O_i\rangle \sim \xi^O$ in $\cal {H}_O$ 
even without  the account of E decoherent interaction. The 
 decoherence, as argued below, only duplicates this effect. 

 
 The incoming S mixture - $|s_i\rangle$ Gemenge results in
 MS algebraic final state 
  $\varphi_{mix}$ which is equivalent to $ \rho^m_{MS}$ of  (\ref {AA3}); 
%
%
  $O$ restricted state  $\varphi^O_{mix}$ 
is  defined from the relation for $\bar{Q}_O$ :
$$
    \bar{Q}_O=\langle\varphi^O_{mix} ;Q_O\rangle=
    \langle\varphi_{mix};Q_O\rangle=\sum |a_i|^2 q^O_i
$$
which results in the solution  $\varphi^O_{mix}=\sum  |a_i|^2\varphi^O_i$,
where $\varphi^O_i=\xi^O_i$ of (\ref {AA66}).
From the  correspondence
of MS state $\xi^{MS}_i$ and $O$ state $\xi^O_i$ in the individual events,
the restricted algebraic
 state $\varphi^O_{mix}$
 represents the statistical mixture of AIS $\xi^O_i$ 
 described by $O$  ensemble state 
  $W^O_{mix}=\{ \xi^O_i;\,P_i=|a_i|^2;\,i=1,2\}$.
%
 If MS final state is  the pure state $\Psi_{MS}$ of (\ref {AA2}) and
 corresponding MS  algebraic state is
$\varphi^{MS}=\xi^{MS}$,  it results 
 in  the same $\bar{Q}_O$ value.
Therefore its $O$ restricted algebraic state
coincides with the mixed one: $\varphi^O=\varphi^O_{mix}$.
From that one can define what are $O$ individual states induced by
MS pure states $\xi^{MS}$, in particular, whether they  differ from
the obtained $\Omega_O$ extremal points $\xi^O_i$ -
the formal individual states.
Remind that any  physically different states can be  operationally
discriminated by the particular observation procedure,
which puts in correspondence  to this states some parameters values.
For the statistical states it can be their probabilistic
distributions parameters. The individual states $\psi_{a,b}$
are   the eigenstates of some $A,B$ observables  
and their difference  is revealed by their
 eigenvalues $q^{A,B}_{i,j}$ which exist objectively.  In principle,
   this values 
 can be extracted  from the single event per each state
(Mittelstaedt,1998). 
If to suppose  that such correspondence maintained also
for the restricted states, then 
in our case the only $\cal{U}_O$ observable is $Q_O$  and
$\xi^O_{i,j}$ difference reflected by $q^O_{i,j}$ values. 
 If to suppose that some other $\xi^O_a \ne \xi^O_i$ exists,
 then some other observable 
 $Q^e \in \cal{U}_O$, for which $\xi^O_a$ is the eigenstate,
 should exist, such that $\, Q^e \ne F(Q^O)$.
However this assumption is inconsistent
 with the obtained $\cal{U}_O$ structure.
 In particular, Breuer  state $R_O$  analog
$\xi^O_R=\sum |a_i|^2 \xi^O_i$ 
can't be $O$ individual state on $\cal{U}_O$, because it
  isn't $Q_O$ eigenstate and can't be $\Omega_O$ extremal point.
%

To illustrate this reasoning,  consider
the superposition $\xi^{MS}_s=\Psi_{MS}$
 of ({\ref{AA2})  with $a_{1,2}=\frac{1}{\sqrt{2}}$; we denote the
(fuzzy) values of $Q^O$ and $B$ of (\ref {AA5}) as $\tilde{q}^O,\tilde{b}$.
Let's define the conditions to which
 the  restricted $\xi^{MS}_s$ state - $\xi^O_s$ should satisfy.  
In $O'$ RF $\xi^{MS}_s$ is $B$ eigenstate with the eigenvalue $\tilde {b}=1$ 
which represents  IT condition,
and $q^O_1 \le \tilde {q}^O \le q^O_2$ is the spectral condition.
Taken together this conditions indicate that $\tilde{q}^O$ is 
 located within the interval $[q^O_1,q^O_2]$ and is
principally uncertain inside it,
as $\tilde b$ value evidences.
 $\xi^{MS}$ restricted state $\xi^O$ is defined on
 $\cal {U}_O$ $=\{I,Q_O\}$,
therefore the spectral condition
 also holds for $O$. Now because $B \notin \cal{U}_O$
and unavailable for $O$,
IT condition can be dropped.  Without it the spectral condition alone
  means  that $\tilde{q}^O$ is   localized in $[q^O_1,q^O_2]$ interval,
 but can be either uncertain or sharp.
Hence $\xi^O_{1,2}$ states with  the sharp $\tilde{q}^O=q^O_{1,2}$
 satisfy to that only necessary  condition for the restricted states.  
 Meanwhile there are no state $\xi^O$ in $\Omega_O$
with $\tilde{q}^O$ uncertain.
The reason of it, as was explained above,
is that  such $\xi^O$ should be an eigenstate of some IT 
observable   which value, alike $\tilde b$,
 can  demonstrate   $\tilde{q}^O$  uncertainty for $O$.
Yet $\cal{U}_O$ doesn't include any such observable.
 Consequently, $\xi^O_{1,2}$ are the only suitable candidates and 
 the formal
 solution for $\xi^{MS} \rightarrow \xi^O_s$ restriction in the 
individual event $n$   is:
 $$
          \xi^O_s(n)=\xi^O_1.or.\xi^O_2
$$
 and it can be shown to be
the same for any $a_{1,2} \neq 0$. It means that $MS\rightarrow O$
 map is ambigous and
  breaks  $Q_O$ symmetry of $\xi^{MS}$. 
 Note also that an arbitrary $\varphi^O$ admits the unique
decomposition into $\xi^O_i$ set and so can be interpreted as $\xi^O_i$ 
ensemble  with the given probabilities $P'_i$.
Thus, in each event $\xi^O_{1}$ or $\xi^O_2$ can appear at random.
For the pure MS ensemble regarded above 
 the only solution which gives  the correct $\bar{Q}_O$  is  $P'_i=|a_i|^2$ 
, and so $W^O=W^O_{mix}$.
 The obtained results demonstrate
  that for MS ensemble
$\xi^{MS}\rightarrow \xi^O$ restriction map is stochastic 
and results in  the  subjective state collapse  observed by $O$ 
analogously  to  DSF state collapse  described above.
%
  Since $\xi^O_i \in \Omega _R$, 
$\xi^O_s$ can be taken  also as the  possible ansatz for
individual MS  states  restriction on 
   $\cal {U}_R$.
DSF doublet state $\Phi$ components $\phi^D,\phi^I$
 correspond to $\xi^{MS}$, $\xi^O$;
the  algebraic states $\varphi(t),\varphi^O(t)$ describe 
$q^O_i$ statistical distributions analogously to DSF $\eta_D,\eta_I$.
%

 Note that MS individual states $\xi^{MS}$ symmetry
 is larger than the symmetry   of the restricted $O$ states.
In Algebraic QM such symmetry reduction results in
 the phenomena of Spontaneous Symmetry Breaking which leads to 
the randomness of outcomes for the models of  measurements in the infinite
systems (Guilini,1996; Mayburov,1998).
The algebraic self-description  permits to extend
such randomness mechanism on the finite systems.
From  the mathematical point of view the
 Algebraic QM formalism contains the generic structure -
the restricted AIS set $\Omega^I_R$ defined on  the
elementary subalgebra
 $\cal {U}^I_R \in \cal U$ ($ \cal {U}^I_R=\cal{U}_O$ in our model).
  $\Omega^I_R$ extremal points, being treated as the individual
 states,  describe the  collapse of the pure states defined on $\cal U$.  
Hence  in this theory
the quantum state reduction results from the reduction of  a system 
algebra to its associative subalgebra.

If to analyze this results  in the Information-Theoretical framework,
 remind that
 the difference between the pure and
mixed MS states reflected by $B$  of (\ref {AA5}) expectation values
(and other IT observables).
Therefore $O$ possible observation of S pure/mixed
 states difference means
that $O$ can acquire the information on   $B$  expectation value.
 But $B \notin \cal {U}_R$ and isn't correlated with $Q_O$ via
S,$O$ interaction alike $Q$ of S, so S,$O$ correlations
supposedly are unobservable from inside by $O$ (Mittelstaedt,1998).
It corresponds to Wigner conclusion that the perception by $O$
of the superposition
of two contradictive impressions is nonsense
and should be excluded in the consistent theory (Wigner, 1961).
Consequently, it's possible that Information Theory rules
by itself can lead
 to the subjective state collapse in the quantum measurements.

 In practice
it's possible that $O$ effective subalgebra is larger than $\cal U_O$,
 but this case
 demands more complicated calculations which we plan to present
in the forcoming paper. In Algebraic QM 
 the only important condition for the classicality appearance
 is $\cal{U}_O$ associativity but it 
  feasible, in principle, also  for the complex IGUS structures.
 If to consider MS+E system 
and its $\cal {U}_{MS,E}$ algebra, the effective $O$ information subalgebra
will be the same $\cal {U}_O$ considered above. 
Therefore $O$ subalgebra and its states set
properties can't depend directly on the surrounding E properties.
%
%
Despite  of the acknowledged  Algebraic QM achievements,
 its foundations are still
discussed and aren't  settled finally. In particular, 
it's still unclear whether all the algebraic states $\varphi$ 
correspond to the
physical states (Primas,1983). This question is  important by itself 
and  can be essential for our formalism feasibility.
 We admitted  also
 that for  MS arbitrary  $\xi^{MS}$  some   $O$ restricted AIS responds
in any event.
 It agrees with the consideration of the restricted states  
 as the real physical states, 
 however, this assumption needs further clarification.
For the regarded simple MS model Algebraic QM formalism in many aspects
is analogous to Orthomodular Algebra of propositions or Quantum Logics
(Jauch, 1968; Emch, 1972). The possibility of
 its application to the systems self-description and the measurement
from inside demands the additional investigation.   



 \section { Discussion}

in this report the
 information-theoretical restrictions on the quantum measurements
were  studied  in the simple selfdescription  model of IGUS $O$.
Breuer self-description theory shows that
by itself   $O$ inclusion  as the quantum object
into the measurement scheme doesn't result in the
 state collapse appearance (Breuer,1996).
Our considerations indicates that to describe the
 state collapse and in the same time to conserve
Schr$\it\ddot{o}$dinger linear evolution, it's necessary 
 to extend the quantum states set over standard QM Hilbert space.
Such modification proposed in DSF leads to the doublet states $\Phi$ ansatz,
where one of its components $\phi^I$ corresponds to $O$
 subjective information - i.e. $O$ selfdescription.
  Algebraic QM  
presents the additional arguments in favor of this approach,
 in its   formalism $O$ structure
  described by $O$ observables algebra $\cal {U}_O$
which defines the multiplet states set analogous to $\Phi$. 
In Algebraic formalism the stochastic events appearance
 stipulated by  MS individual states restriction to $O$.
 Algebraic QM formalism
-  Segal and $C^*$-algebras of operators is acknowledged
generalization of standard QM (Emch, 1972).
In addition, from the mathematical point of view 
the duality of the operators algebra and the states set
is better founded, than the states set priority
postulated in standard QM (Bratelli, 1981).
 In this paper it was applied for 
the simplest measurement model but if this formalism universality
will be proved, it would mean that the proposed measurement
 theory follows from the established Quantum Physics realm (Emch,1972).
 Algebraic formalism  permits to calculate
  the restricted individual $O$ states $\xi^O$  ansatz
 without any phenomenological assumptions.
From the formal point of view the only novel feature of our approach 
is the use of Segal algebra for the individual restricted
 states $R_O(n)$ calculations.
We don't see any compelling mathematical arguments why  
 for the individual states $R_O$ 
 ansatz should be chosen  the same QM formulae (\ref {AA4})   
  which used for  the statistical restricted states
$R^{st}_O$ (Lahti,1990).



Our  theory demonstrates that the probabilistic realization
 is generic and unavoidable for QM and without it QM supposedly can't
acquire any operational meaning. Wave-particle dualism
was always regarded as characteristic QM  feature but in our theory
it has straightforward  correspondence in our  DSF.
 Our  approach stresses also the
 dual character of
quantum measurement : this is the interaction 
of studied S with IGUS $O$ and in the same
time the information acquisition and recognition by IGUS. 
  Note that Self-reference problem
 avoided in this case by use of the natural 
 assumption that all observers are similar in relation to
 their information acquisition properties (Svozil,1993).


Under the realistic conditions, the rate of E atoms interactions with 
macroscopic detector D is very high, and due to it in a very short time $t_d$
S,D partial state $\rho_p=Tr_E \rho_{SDE}$ becomes approximately equal
to the mixed one, as $\rho_p$ nondiagonal elements become
very small. This fact induced the claim
that the objective state collapse   can be completely explained by detector  
state decoherence without the observer's inclusion, but it
was proved to be incorrect (D'Espagnat,1990).
The development of decoherence approach proposed
  in Zurek 'Existential interpretation (Zurek,1998).
 IGUS   $O$  regarded as the  quantum object and
 included in the measurement chain; 
  memorization of input S signal  
occurs in several binary memory cells $|m^j_{1,2}\rangle$ (chain)
which are the analog of  the brain neurons. 
 $O$ memory state suffers  the decoherence from surrounding E 'atoms'
which results in the system state analogous to (\ref {DD1}).  
Under  practical  conditions, the decoherence time $t_d$ 
is also  small and for $t \gg t_d$
 S,$O$ partial state $\rho_p$ differs from the mixture
very little.  From that  Zurek concludes that $O$ 
percepts input pure S signal as the random measurement outcomes.
Yet the system S,$O$,E 
 is still in the pure state even at $t\gg t_d$ and  IT observable
$B$ analogous to (\ref {AA5})
  which proves it exists.  Therefore in standard QM framework
it's incorrect to claim that IGUS percepts random events. 
The regarded IGUS model doesn't differs principally
from our MS scheme;
 in Algebraic formalism IGUS  subjective perception  is described
by  $O$ restricted  state $\xi^O$ defined on $\cal {U}_O$
 which describes the random
outcomes for the input pure S state.  Consequently, 
Algebraic QM  application 
  to Zurek IGUS model supports Existential Interpretation.

It's important to stress that all the   experiments in Physics  at the
final stage include  the human subjective perception
which simulated by $O$ state in our model.
The possible importance of observer in the measurement
process was discussed first by London and Bauer (London,1939). They supposed
that  Observer Consciousness (OC) due to 'introspection
action' violates in fact Schrodinger equation  
and results in the state reduction.
 In distinction, in our  theory $O$ perception     
 doesn't violate MS Schrodinger evolution from $O'$ point of view.
In principle,
Self-description Theory  permits  to regard the relation between
MS, IGUS $O$ states and $O$ subjective information (impression), and
 one can try to extend it on the human perception.
 This is the separate, important problem which is beyond
our scope and here
we consider briefly only some its principal points which 
formulated here as the simple Impression Model (IM). 
Following our approach, $O$ perception affected only by 
$O$ internal states defined on $O$ observables subalgebra $\cal {U}_O$. 
For  $O$ perception   the following, calibration assumption
introduced:  for any Q eigenstate $|s_i\rangle$ 
 after S measurement finished at $t>t_1$ 
and $O$ 'internal pointer'state is $|O_i\rangle$ observer $O$
 have the definite impression $I^O$ 
corresponding  to $I^O=q^O_i$ eigenvalue  - the information 
pattern percepted by $O$.
It  settles the hypothetical correspondence between MS quantum
dynamics model and human perception. 
Impression $I^O$ is $O$ subjective information which isn't
dynamical parameter and  its introduction can't have
any influence on  the theory dynamics.
 Furthermore, we assume that
  if S state is the superposition $\psi_s$  
 then its measurement by $O$ also  results in appearance
for each individual event $n$ of some 
 definite  and unambiguous $O$ impression  $I^O=q^{sup}(n)$. 
 In Algebraic formalism the corresponding
  $O$ subjective information - impression in the individual
event  is equal to  $I^O(n)=q^O_i$ and can be consistently defined
in this ansatz for our IM as the stochastic state 
appearing  with probability $|a_i|^2$.
In algebraic QM  framework they correspond to the restricted $O$ AIS $\xi^O_i$
defined on $\cal{U}_O$.
The obtained picture is the analog
 of Von Neuman psychophisical parallelism hypothesis.
Yet we must  stress that, on the whole,  our  theory doesn't need any
addressing to  human OC. Rather, in this model IGUS
$O$ is active RF which internal state excited by the interaction
with the studied object.

To conclude, the quantum measurements were studied within the
 Information-Theoretical framework and  self-description
 restrictions  on the information acquisition
are shown to be important in the Measurement Theory.
Algebraic QM represents
the appropriate formalism of systems self-description,  
 in particular, 
 IGUS  $O$ observable algebra $\cal {U}_O$ defines $O$ restricted states
$\xi^O$ set $\Omega_O$.
The appearance of stochastic events 
 stipulated by  MS individual states restriction to  $\xi^O$ states
and results in the  state collapse observation by $O$. 
The regarded IGUS model is quite simple and on the whole doesn't
permit us to make any final conclusions at this stage.
Yet  the obtained results evidence
 that the  IGUS information
 restrictions and its interactions with the observed system 
should be accounted in Quantum  Measurement Problem analysis  (Zurek,1998).
\\
\\
\qquad \qquad \qquad {\Large { References}}\\
\\
 (1981) Y.Aharonov, D.Z. Albert Phys. Rev. D24, 359 
\\
 (2000) G.Bene, quant-ph 0008128
\\
 (1979) O.Bratteli, D.Robinson 'Operators Algebra and
Quantum Statistical Mechanics' (Springer-Verlag, Berlin)
\\
 (1996) T.Breuer, Phyl. of Science 62, 197 (1995),
 Synthese 107, 1 (1996)
\\
 (1996) P.Busch, P.Lahti, P.Mittelstaedt,
'Quantum Theory of Measurements' (Springer-Verlag, Berlin,1996)
\\
 (1990) W. D'Espagnat, Found Phys. 20,1157,(1990)
\\
%
 (2002) R.Duvenhage, Found. Phys. 32, 1799  
\\
 (1994) A.Elby, J.Bub Phys. Rev. A49, 4213 
\\
 (1972) G.Emch, 'Algebraic Methods in Statistical Physics and
Quantum Mechanics',\\
 (Wiley,N-Y) 
\\
 (1988) D.Finkelstein, 'The Univrsal Turing Machine:
 A Half Century Survey', (ed. R.Herken, University Press, Oxford) 
\\
 (1996) D.Guilini et al., 'Decoherence and Appearance of
Classical World', (Springer-Verlag,Berlin) 
\\
 (1985) S.Kochen 'Symposium on Foundations of Modern Physics'
  , (World scientific, Singapour)
\\
(1968) J.M.Jauch 'Foundations of Quantum Mechanics' 
(Adison-Wesly, Reading)
\\
 (1990) P. Lahti Int. J. Theor. Phys. 29, 339 
\\
 (1939)  London F., Bauer E. 'La theorie de l'Observation'
 (Hermann, Paris)   
\\
 (1998) S.Mayburov, Int. Journ. Theor. Phys. 37, 401 
\\
 (2001) S.Mayburov  Proc. V  QMCC Conference, Capri, 2000,
(Kluwer, N-Y); $ \setminus$quant-ph 0103161
\\
 (2002) S.Mayburov Proc. of Vth Quantum Structures conference,
Cesenatico, 2002; Quant-ph 0205024; Quant-ph 0212099
\\
 (1998) P.Mittelstaedt 'Interpretation of
Quantum Mechanics and Quantum Measurement\\ Problem',
(Oxford Press, Oxford)
\\
 (1932) J. von Neuman 'Matematische Grunlanden
 der Quantenmechaniks' , (Berlin)
\\
 (1983) H.Primas,  'Quantum Mechanics,
 Chemistry and Reductionism' (Springer, Berlin)
 (1990) H.Primas,  in  'Sixty two years of uncertainty'
,ed. E.Muller, (Plenum, N-Y)
\\
 (1995) C. Rovelli, Int. Journ. Theor. Phys. 35, 1637; 
quant-ph 9609002  
\\
 (1947) I.Segal, Ann. Math., 48, 930    
\\
 (1993) K.Svozil 'Randomness and undecidability in Physics',
(World Scientific, Singapour)
\\
%
 (1961) E.Wigner,  'Scientist speculates',(Heinemann, London)
\\
 (1982) W.Zurek, Phys Rev, D26,1862 
\\
 (1998) W.Zurek Phys. Scripta , T76 , 186 
\\
\\
\end {document}